# Inverse design of multilayer nanoparticles using artificial neural networks and genetic algorithm

Cankun Qiu, Zhi Luo, Xia Wu, Huidong Yang, and Bo Huang*

College of Information Science and Technology, Jinan University, Guangzhou 510632, China

## Abstract

The light scattering of multilayer nanoparticles can be solved by Maxwell's equations. However, it is difficult to solve the inverse design of multilayer nanoparticles by using the traditional trial-and-error method. Here, we present a method for forward simulation and inverse design of multilayer nanoparticles. We combine the global search ability of genetic algorithm with the local search ability of neural network. First, the genetic algorithm is used to find a suitable solution, and then the neural network is used to fine-tune it. Due to the non-unique relationship between physical structures and optical responses, we first train a forward neural network (structure-to-spectrum), and then it is applied to the inverse design of multilayer nanoparticles. Not only here, this method can easily be extended to predict and find the best design parameters for other optical structures.

## Introduction

Nanoparticles have many typical properties and one of the most attractive aspects is that their optical property can be manipulated by adjusting their internal structure. Just as the metallic nanoparticles have many interesting optical properties due to the localized surface plasmon resonance (LSPR), whose resonance frequency can be managed by using different combinations of shell thickness and materials[1]. Nowadays, the research of nanoparticles has attracted extensive attention because of their unusual optical application in biological imaging[2]–[4] and detection[5], photothermal therapy[6], [7], catalytic processes[8] and nonlinear optics[9]. As nanoparticles are increasingly used in various fields and a great amount of progress has been made in the synthesis of nanoparticles[10], [11], it is urgent to find a suitable design method for the inverse design of nanoparticles. Conventional design approaches used in the inverse design can be divided into two groups: the evolutionary method[12]–[15], and the gradient-based method[16], [17]. Here, we adopt deep learning[18] and genetic algorithm to the inverse design of multilayer nanoparticles, avoiding conventional design approaches based on trial-and-error which is computationally expensive. Neural networks (NNs) have been shown to be very effective in approximating many optical simulations and inverse designs[19]–[26]. Recently, Ali Adibi et. al. used NNs based on autoencoder for designing electromagnetic nanostructure[27], Kun Xu et. al. used NNs to inverse design of the plasmonic waveguide coupled with cavities structure[28], and Jinfeng Zhu used NNs to inverse design of graphene-based photonic metamaterials[29]. In photonics, the optical scattering of nanoparticles can be well understood by Maxwell's equations. For this purpose, there are many tools have been developed for forward simulation, such as the finite-difference time-domain (FDTD)[30], finite element method (FEM)[31], transfer matrix method[32], [33] and so on. Compared with traditional electromagnetic simulation method, neural networks can reduce a lot of time used to approximate the EM simulation.

As a data-driven approach, deep learning requires a large amount of data to train a forward neural network and it takes a lot of time to obtain these data using simulation, especially when simulate multi-layer nanoparticles. But this is a one-time consumption, it can predict the spectrum faster than simulation after the forward network was trained. In order to reduce the value of loss function of the network, we always need to adjust the hyperparameters or increase the complexity of the network and the amount of data. Here, we adopt a two-channel neural networks (TCNNs) to reduce the loss between predicted and exact spectrum and enhance the robustness of the network as shown in Fig. 1(b). A good forward neural network is useful for the inverse design of optics. This not only makes the spectrum of inverse design more similar to the exact spectrum, but also avoids the neural network oscillating in the structure where the predicted spectrums have similar sum of the squared errors with the exact spectrum. Experiments show that the average validation error of each spectrum is reduced by twice as much as that of the conventional fully connected neural network when using the two-channel neural network. In the inverse design of multilayer nanoparticles, training the inputs of neural network[21] or using the tandem neural network[22] is a good way when the input dimension is low, but the performance of neural network is poor and neural network is difficult to jump out of local minimum or saddle point when the input dimension is high. In this paper, we use TCNNs and genetic algorithm to effectively solve the inverse design of multilayer nanoparticles.

## 1. Training a forward neural network

Here, we first train a forward neural network that it can approximate the Maxwell equations. Compared with traditional electromagnetic simulation, the method based on NNs can be simulate more quickly after the NNs is trained. In the multilayer nanoparticles, the spectrum has more sharp peaks in the shorter wavelength range when each layer is between 30 and 70 nanometers. Besides, experiments show that neural networks tend to perform well when the input dimension (this refers to the number of layers of nanoparticles) is low. When the input dimension is increasing, the output spectrum is often steeper than that of the input lower dimension, and the performance of the neural network is worse. This can be improved by adding data, but adding data is time consuming by EM simulation. Here, we adopt two-channel neural networks that the contribution of the error of different wavelength range to the total training error is different (see formula (1)), and the results show that the TCNNs as a forward neural network has lower validation error under the same data comparing conventional fully connected neural networks (FCNNs) that the output of each layer of neurons is sent to all the nodes of the next layer.

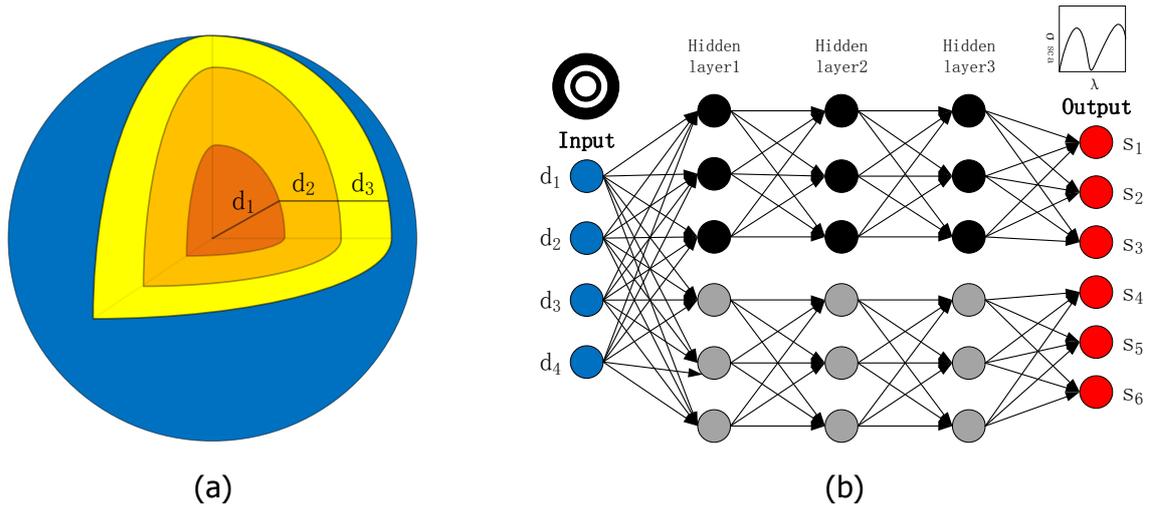

Figure 1: (a) Schematics of a nanoparticle. (b) A forward neural network. The neural network takes the thickness of each layer of the nanoparticle as inputs and outputs scattering cross section at different wavelengths of the scattering spectrum. The neural network consists of two sub-networks, the upper sub-network predicts the first half of the spectrum, and the lower sub-network predicts the second half of the spectrum. The actual subnetworks have seven hidden layers, each with 250 neurons.

We used a TCNNs which consists of two fully connected subnetworks. As shown in Fig. 1(b), we consider a particle consisting alternately of $SiO_2$ and $TiO_2$, and the neural network inputs the thickness of nanoparticles D and outputs the scattering cross section at different wavelengths of the scattering spectrum S. we generate 20,0000 data using the transfer matrix method[33]. 190,000 data were used for training and the rest for verification and test sets. The activation function of neural networks are "scaled exponential linear units" (SELUs), which induce self-normalizing properties[34]. Adam optimizer[35] is used to update the weights of the network. The weights and bias of two subnetworks are updated simultaneously by back-propagation[36]. And the weights and bias of the network will be fixed after 1000 iterations. The errors in training are shown in Fig. 2(a), and we use the SSE (sum of the squared errors) as the cost function. The training loss is defined by

$$\text{Loss} = m \times \sum_{i=0}^{\frac{n}{2}-1} (S_p(\lambda_i) - S_e(\lambda_i))^2 + (1-m) \times \sum_{i=\frac{n}{2}}^{n-1} (S_p(\lambda_i) - S_e(\lambda_i))^2 \quad (1)$$

And the validation error is given by
$$Error = \sum_{i=0}^{n-1} (S_p(\lambda_i) - S_e(\lambda_i))^2 \quad (2)$$

Except for special instructions in the following, the verification error is used as the cost function except the training of TCNNs which use the training loss. where n is the number of spectral sampling points, $S_p(\lambda_i)$ and $S_e(\lambda_i)$ are the predicted and the exact results of each spectral point respectively. In this paper, the output is 400 spectrum points between 400 and 800 nanometers, so n=400. And m is a hyperparameter between 0 and 1. Here, m equal to 0.6 is reasonable after lots of experiments.

We compare the TCNNs with the FCNNs where the number of neurons in each hidden layer is 520 and the other conditions are the same as the two-channel neural networks. The average validation error of two-channel neural network is only half that of conventional fully connected neural network, as shown in Fig. 3. Next, we test an example out of training set and find that forward neural network can approximate spectrum it was not train on well as shown in Fig. 2(b). In fact, it doesn't take much time to train such a network based on 2 GTX 1070 graphics card. Once the neural network is trained, it can easily predict the spectrum than traditional electromagnetic simulation.

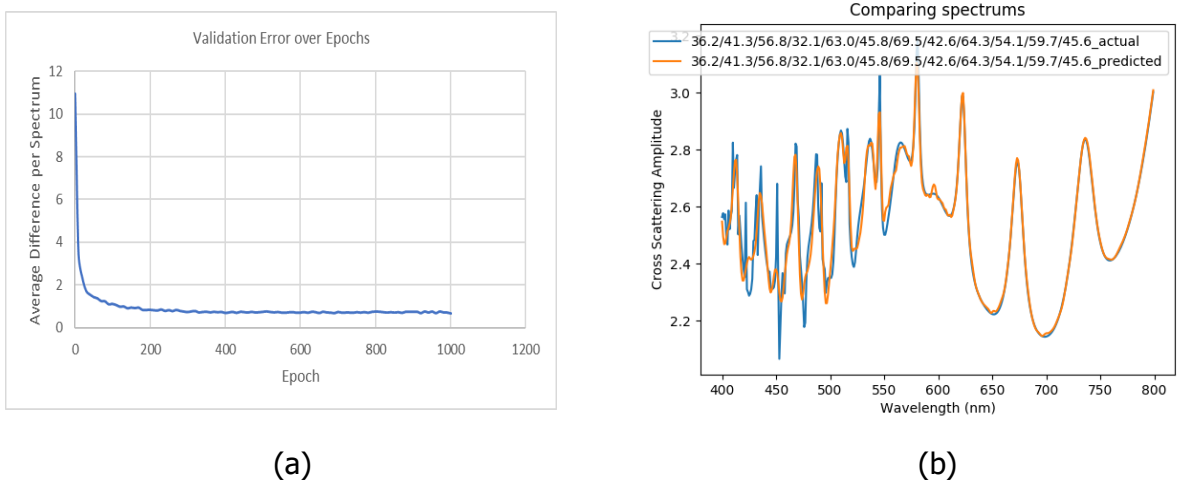

(a)          (b)

Figure 2: (a) The mean error of spectrum of validation set for the twelve layers case. It can be seen that the network declines rapidly at the first 100 epochs, and gradually converged after 200 epochs. (b) After the network training, an example which was not used for neural network training is randomly selected from the test set for testing. The blue line is desired spectrum, and the yellow line is the output of the neural network. It can be seen from the figure that the output results of neural network are in good agreement with the exact results of simulation. These results were consistent across many different spectra.

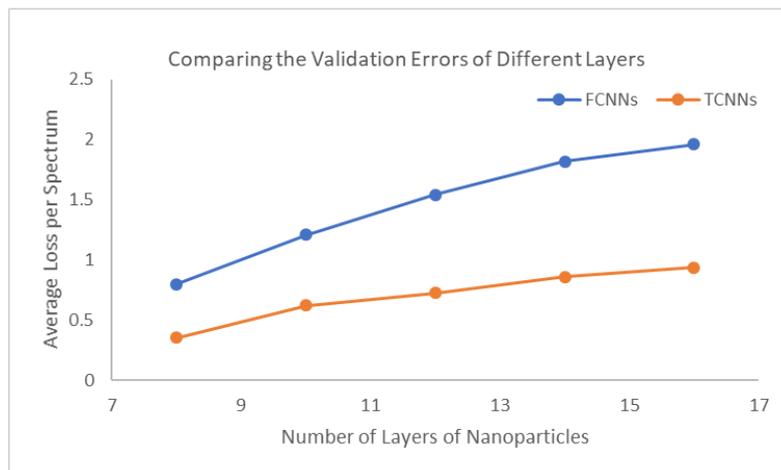

Figure 3: Comparing the average validation errors of different layers between TCNNs and FCNNs. FCNNs become unfitted as the number of layers of nanoparticles increases. However, the loss of TCNNs increases more slowly.

## 2. Inverse design of multilayer nanophotonics

Due to the inconsistency of data, the structure of multiple nanoparticles produces the same spectrum. It is difficult to converge that train a network that input the structural parameters of nanoparticles and output the spectra. Now, many methods[21]–[23], [27], [37] have been proposed to solve inverse design problems. However, the inverse design of multilayer nanoparticles is a non-convex optimization problem in a high dimensional space such as there are a lot of saddle points and extreme points, making it very difficult to find a global optimal solution. In this paper, we combine genetic algorithm (GA) and TCNNs. GA is a random search algorithm that simulates the evolutionary process of an organism, and it's good with searching for a solution in the vicinity of the optimal solution. It performs a series of genetic operations such as selection, crossover and mutation on the population to produce better generations. When the design parameters are near the optimal solution, the neural network based on back-propagation can quickly converge to the optimal solution.

We demonstrated how to design a 12-layer nanoparticle using GA and TCNNs, each layer between 30 and 70 nanometers. We limit the thickness of each layer of nanoparticles to 35,45,55 or 65 in the initial population. If the thickness of nanoparticles in the initial population take a random number between 30 and 70 nanometers, the search space will become large, and it is very time-consuming for genetic algorithm to search for an optimal solution. In other words, the genetic algorithm does a preprocessing instead of searching the optimal solution, and it outputs a sequence of Numbers that are close to the optimal solution, consisting of 35,45,55 and 65. Of course, we can't use a full search to traverse every possible design parameter either. It works for only a few layers of nanoparticles, but the solution space for multiple layers of nanoparticles would be huge, such as 16 million orders of magnitude for 12-layer nanoparticles, making it nearly impossible to use full search. The flowchart of GA is shown in Fig. 4. The number of the initial population is set as 100. Here roulette wheel selection[38] as a selection method of GA. and the number of selection:

$$N_{selection} = \frac{Max(X)}{T_{value}} \otimes Mean(X) \qquad (3)$$

In formula (3), X = [X1, X2, …, X100], Where $X_i$ is the fitness function value of each individual in the population calculated by the forward neural network:

$$X_i = \frac{1000n}{\sum_{j=1}^{n}(S_p(\lambda_j)-S_e(\lambda_j))^2} \quad (i=1,2,3…100) \qquad (4)$$

In addition, the number of crossover and mutation are

$$N_{crossover} = 0.7 \otimes (100 - N_{selection}) \qquad (5)$$

and

$$N_{mutation} = 100 - N_{selection} - N_{crossover} \qquad (6)$$

respectively. If N selection is greater than or equal to 90, N selection equal to 90. Using this method of population selection can help us adjust the number of selected individuals reasonably so that the GA can reach the specified fitness function value faster. With adaptive selection operation, there is a higher probability of crossover and mutation at the beginning, and more new individuals will be created. When the maximum fitness and average fitness of the population are increasing, the probability of selection operation of GA is increasing, and the probability of crossover and mutation is decreasing, so more individuals from last population will be retained. As shown in figure 5, the maximum fitness of the adaptive selection operation increases rapidly at the beginning, which is equivalent to giving a small selection probability. When an individual with relatively large fitness is found, the probability of selection will increase, and the probability of survival of the individual with the maximum fitness increases. Even if the maximum adaptive fitness of adaptive selection operation has a flat period, when the average adaptive fitness rises to a certain value, the maximum adaptive fitness will continue to rise. However, when the probability of the selection operation is fixed at $N_{selection}=20$, the maximum fitness individual is easily discarded as the dotted circle shown in the figure, even though the fitness increases rapidly at the beginning. And the maximum fitness is difficult to increase at the beginning when the probability of the selection operation is fixed at $N_{selection}=80$, which is also unreasonable when the initial population is still at a low fitness level. Even though there are exceptions, the selection operation of $N_{selection}=20$ does not lose the maximum fitness individual, and the initial population of the selection operation of $N_{selection}=80$ has a good overall fitness, but in general, the adaptive selection operation is relatively stable. Therefore, the adaptive selection operation would find an individual with large fitness faster without throwing it away. Finally, when the fitness of the optimal individual reaches the given threshold or the number of iterations reaches the preset value, the algorithm terminates.

After preprocessed by GA, the best individual input into the neural network that the weights and bias remain unchanged trained by the forward neural network, and the input of the neural network is fine-tuned by the back-propagation. Through several experiments, we found that the neural network can converge to a minimum value even though the optimal fitness and individual output of genetic algorithm may be different. It shows that the neural network can solve the inverse design of nanoparticles well after the data is preprocessed by genetic algorithm. Experimental results are shown in Fig. 6(a). We also compare the method proposed in this paper with the other two methods, training the input of neural network directly[21] and inverse design with a tandem network[22]. These methods perform well in low-dimensional design space. But in high-dimensional design space, where the latter two methods need a large amount of time to adjust the hyperparameters of neural network, the method proposed in this paper is obviously more effective than the other two methods.

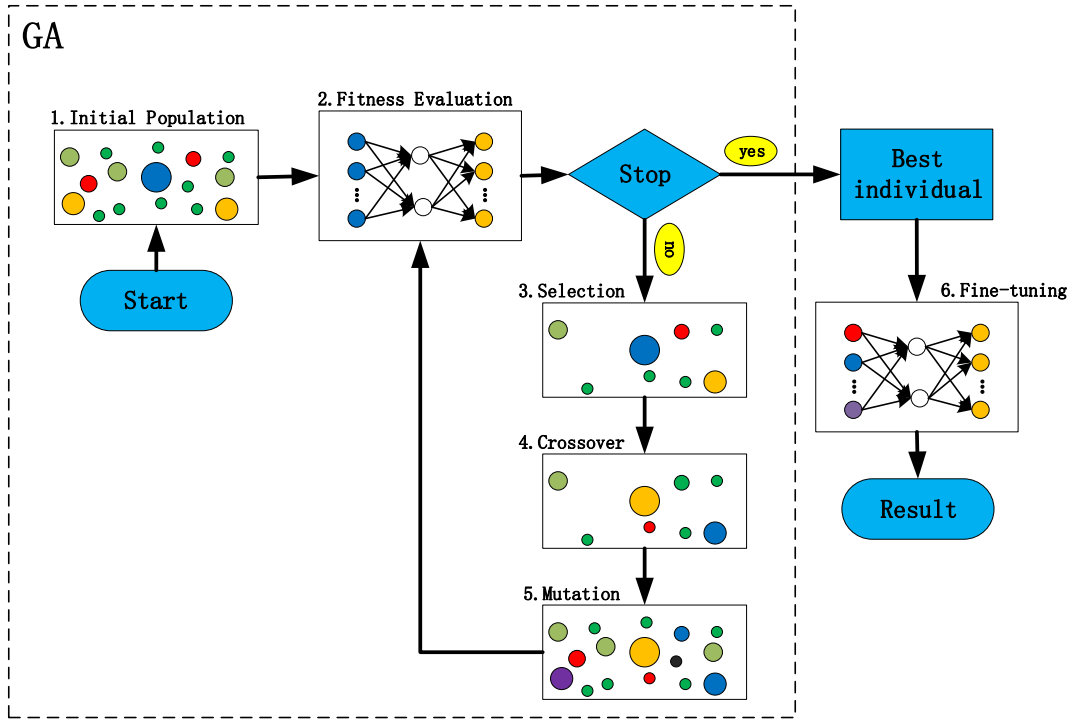

Figure 4: The flowchart of inverse design. The initial population was randomly generated and fitness was calculated by forward neural network. When the fitness accord with a given threshold or the number of iterations reaches a given value, the iteration ends. Otherwise, populations are optimized generation by generation through selection, crossover, and mutation. At the end of the iteration, the result of GA is sent to the neural network for fine-tuning.

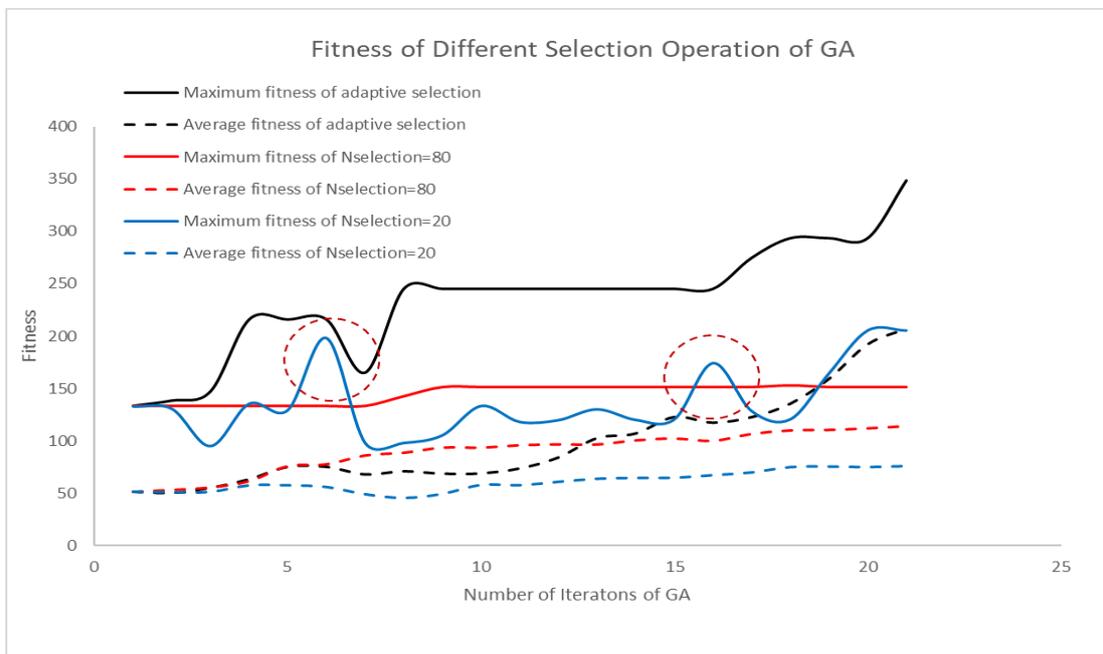

Figure 5: An example of the comparison of adaptive selection operation with the selection operations of $N_{selection}=20$ and $N_{selection}=80$, which has the same initial population.

The solid line is the maximum fitness change curve and the dotted line is the average fitness change curve (The fitness calculation is shown in formula 4). Compared with the selection operations N $_{selection}$ =20 and N $_{selection}$ =80, the adaptive selection operation would find an individual with high fitness faster. This reduces the irrationality of selecting a large number of individuals with low fitness and discarding the maximum fitness in the iteration of GA due to the fixed N $_{selection}$.

We also apply the above method to higher dimensional data and find it very useful. Fig. 6(b). shows the design of a 16-layer nanoparticle. Although there are many sharp peaks in the spectrum, we can still use the above method for the inverse design of nanoparticles. Therefore, this shows that our method is useful in dealing with higher dimensional problems.

(a)

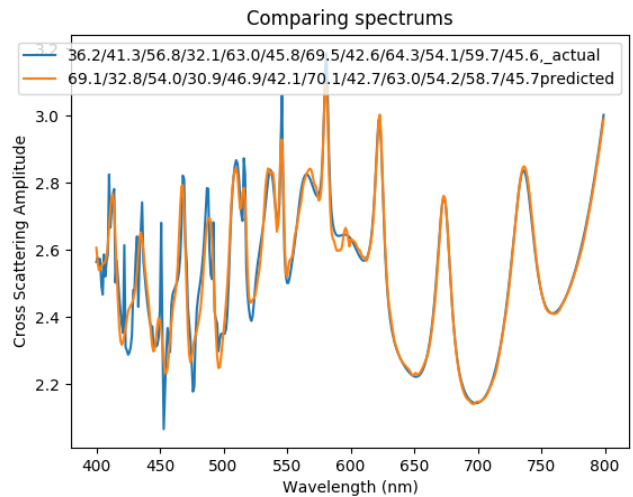

(b)

Example 1:

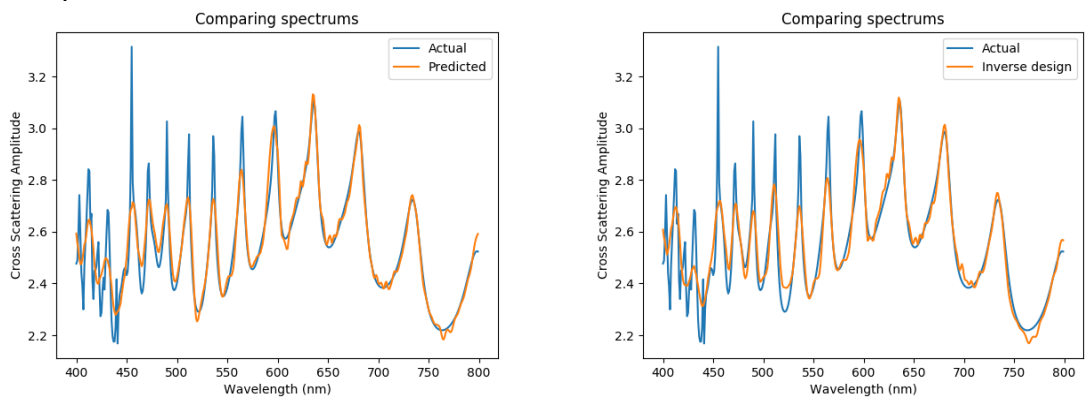

FIG. 6: (a) Compare the exact spectrum with the spectrum produced by the predicted structure. It is shown that the network can generate the thickness of the multilayer nanostructure that conforms to the desired spectrum. The blue is the exact spectrum, and the yellow is the predicted spectrum produced by the predicted nanostructure. (b) The example of 16-layer nanoparticles.

## 3. Conclution

Here, we propose to use the TCNNs in the forward neural network and GA with adaptive selection operation in the inverse design. We demonstrate that using the TCNNs can effectively improve the accuracy of prediction in the same dataset, and it's helpful for NNs with small datasets. For inverse design, training the input of NNs which trained by a forward neural network is easy to fall into saddle point or local minimum value in high dimensional design space, and the neural network is difficult to jump out. Using genetic algorithm as a filter to pick out some good initial values which effectively improve the processing capacity of NNs. The method proposed in this paper provides a way to design the inverse design of complex photonic structures.

## References

[1] E. Hutter and J. H. Fendler, "Exploitation of Localized Surface Plasmon Resonance," *Adv. Mater.*, vol. 16, no. 19, pp. 1685–1706, Oct. 2004.
[2] P. K. Jain, K. S. Lee, I. H. El-Sayed, and M. A. El-Sayed, "Calculated Absorption and Scattering Properties of Gold Nanoparticles of Different Size, Shape, and Composition: Applications in Biological Imaging and Biomedicine," *J. Phys. Chem. B*, vol. 110, no. 14, pp. 7238–7248, Apr. 2006.
[3] J. Yguerabide and E. E. Yguerabide, "Light-Scattering Submicroscopic Particles as Highly Fluorescent Analogs and Their Use as Tracer Labels in Clinical and Biological Applications: II. Experimental Characterization," *Analytical Biochemistry*, vol. 262, no. 2, pp. 157–176, Sep. 1998.
[4] M. Liong *et al.*, "Multifunctional Inorganic Nanoparticles for Imaging, Targeting, and Drug Delivery," *ACS Nano*, vol. 2, no. 5, pp. 889–896, May 2008.
[5] R. Elghanian, J. J. Storhoff, R. C. Mucic, R. L. Letsinger, and C. A. Mirkin, "Selective Colorimetric Detection of Polynucleotides Based on the Distance-Dependent Optical Properties of Gold Nanoparticles," *Science*, vol. 277, no. 5329, pp. 1078–1081, Aug. 1997.
[6] "Photothermal therapy in a murine colon cancer model using near-infrared absorbing gold nanorods." [Online]. Available: https://www.spiedigitallibrary.org/journals/Journal-of-Biomedical-Optics/volume-15/issue-1/018001/Photothermal-therapy-in-a-murine-colon-cancer-model-using-near/10.1117/1.3290817.full. [Accessed: 10-Aug-2019].
[7] X. Huang, I. H. El-Sayed, W. Qian, and M. A. El-Sayed, "Cancer Cell Imaging and Photothermal Therapy in the Near-Infrared Region by Using Gold Nanorods," *J. Am. Chem. Soc.*, vol. 128, no. 6, pp. 2115–2120, Feb. 2006.
[8] W. Tang and G. Henkelman, "Charge redistribution in core-shell nanoparticles to promote oxygen reduction," *J. Chem. Phys.*, vol. 130, no. 19, p. 194504, May 2009.
[9] Y. Pu, R. Grange, C.-L. Hsieh, and D. Psaltis, "Nonlinear Optical Properties of Core-Shell Nanocavities for Enhanced Second-Harmonic Generation," *Phys. Rev. Lett.*, vol. 104, no. 20, p. 207402, May 2010.
[10] S. Hasan, "A Review on Nanoparticles: Their Synthesis and Types," vol. 4, p. 4, 2015.
[11] M. T. Swihart, "Vapor-phase synthesis of nanoparticles," *Current Opinion in Colloid & Interface Science*, vol. 8, no. 1, pp. 127–133, Mar. 2003.

[12] J. H. Holland, P. of P. and of E. E. and C. S. J. H. Holland, and S. L. in H. R. M. Holland, *Adaptation in Natural and Artificial Systems: An Introductory Analysis with Applications to Biology, Control, and Artificial Intelligence*. MIT Press, 1992.

[13] "Genetic Algorithms and Machine Learning," p. 5.

[14] R. Eberhart and J. Kennedy, "A new optimizer using particle swarm theory," in *MHS'95. Proceedings of the Sixth International Symposium on Micro Machine and Human Science*, 1995, pp. 39–43.

[15] J. Robinson and Y. Rahmat-Samii, "Particle swarm optimization in electromagnetics," *IEEE Transactions on Antennas and Propagation*, vol. 52, no. 2, pp. 397–407, Feb. 2004.

[16] M. P. Bendsøe and N. Kikuchi, "Generating optimal topologies in structural design using a homogenization method," *Computer Methods in Applied Mechanics and Engineering*, vol. 71, no. 2, pp. 197–224, Nov. 1988.

[17] M. P. Bendsoe and O. Sigmund, *Topology Optimization: Theory, Methods, and Applications*, 2nd ed. Berlin Heidelberg: Springer-Verlag, 2004.

[18] Y. LeCun, Y. Bengio, and G. Hinton, "Deep learning," *Nature*, vol. 521, no. 7553, pp. 436–444, May 2015.

[19] M. M. Vai, Shuichi Wu, Bin Li, and S. Prasad, "Reverse modeling of microwave circuits with bidirectional neural network models," *IEEE Trans. Microwave Theory Techn.*, vol. 46, no. 10, pp. 1492–1494, Oct. 1998.

[20] H. Kabir, Ying Wang, Ming Yu, and Qi-Jun Zhang, "Neural Network Inverse Modeling and Applications to Microwave Filter Design," *IEEE Trans. Microwave Theory Techn.*, vol. 56, no. 4, pp. 867–879, Apr. 2008.

[21] J. Peurifoy *et al.*, "Nanophotonic Particle Simulation and Inverse Design Using Artificial Neural Networks," *arXiv:1712.03222 [physics]*, Oct. 2017.

[22] D. Liu, Y. Tan, E. Khoram, and Z. Yu, "Training Deep Neural Networks for the Inverse Design of Nanophotonic Structures," *ACS Photonics*, vol. 5, no. 4, pp. 1365–1369, Apr. 2018.

[23] Z. Liu, D. Zhu, S. P. Rodrigues, K.-T. Lee, and W. Cai, "Generative Model for the Inverse Design of Metasurfaces," *Nano Lett.*, vol. 18, no. 10, pp. 6570–6576, Oct. 2018.

[24] S. So and J. Rho, "Designing nanophotonic structures using conditional-deep convolutional generative adversarial networks," *arXiv:1903.08432 [physics]*, Mar. 2019.

[25] S. Inampudi and H. Mosallaei, "Neural network based design of metagratings," *Appl. Phys. Lett.*, vol. 112, no. 24, p. 241102, Jun. 2018.

[26] S. An *et al.*, "A Novel Modeling Approach for All-Dielectric Metasurfaces Using Deep Neural Networks," p. 18.

[27] Y. Kiarashinejad, S. Abdollahramezani, and A. Adibi, "Deep learning approach based on dimensionality reduction for designing electromagnetic nanostructures," *arXiv:1902.03865 [physics, stat]*, Feb. 2019.

[28] T. Zhang *et al.*, "Efficient spectrum prediction and inverse design for plasmonic waveguide systems based on artificial neural networks," *Photon. Res.*, vol. 7, no. 3, p. 368, Mar. 2019.

[29] Y. Chen, J. Zhu, Y. Xie, N. Feng, and Q. Huo Liu, "Smart inverse design of graphene-based photonic metamaterials by an adaptive artificial neural network," *Nanoscale*, vol. 11, no. 19, pp. 9749–9755, 2019.

[30] A. F. Oskooi, D. Roundy, M. Ibanescu, P. Bermel, J. D. Joannopoulos, and S. G. Johnson, "Meep: A flexible free-software package for electromagnetic simulations by the FDTD method," *Computer Physics Communications*, vol. 181, no. 3, pp. 687–702, Mar. 2010.

[31] J.-M. Jin, *The Finite Element Method in Electromagnetics*. Wiley, 2015.

[32] "Bohren C.F._ Huffman D.R. Absorption and scattering of light by small particles.pdf." .
[33] W. Qiu, B. G. DeLacy, S. G. Johnson, J. D. Joannopoulos, and M. Soljačić, "Optimization of broadband optical response of multilayer nanospheres," *Opt. Express*, vol. 20, no. 16, p. 18494, Jul. 2012.
[34] G. Klambauer, T. Unterthiner, A. Mayr, and S. Hochreiter, "Self-Normalizing Neural Networks," *arXiv:1706.02515 [cs, stat]*, Jun. 2017.
[35] D. P. Kingma and J. Ba, "Adam: A Method for Stochastic Optimization," *arXiv:1412.6980 [cs]*, Dec. 2014.
[36] "A-Theoretical-Framework-for-Back-Propagation.pdf." .
[37] W. Ma, F. Cheng, Y. Xu, Q. Wen, and Y. Liu, "Probabilistic representation and inverse design of metamaterials based on a deep generative model with semi-supervised learning strategy," *arXiv:1901.10819 [physics]*, Jan. 2019.
[38] A. Lipowski and D. Lipowska, "Roulette-wheel selection via stochastic acceptance," *Physica A: Statistical Mechanics and its Applications*, vol. 391, no. 6, pp. 2193–2196, Mar. 2012.